\documentclass[aps, prd, showpacs, showkeys, preprint]{revtex4}
\begin{document}
\begin{titlepage}
~\vskip1cm
\title{ Comment on "Infrared freezing of Euclidean QCD observables"}
\author{Irinel Caprini\footnote{caprini@theory.nipne.ro}}
\affiliation{National Institute of Physics and Nuclear Engineering,Bucharest POB MG-6, R-077125 Romania}
\author{Jan Fischer\footnote{fischer@fzu.cz}}
\affiliation{Institute of Physics, Academy of Sciences of the Czech Republic,CZ-182 21  Prague 8, Czech Republic}

\begin{abstract}  Recently, P. M. Brooks and C.J. Maxwell 
 [Phys. Rev. D{\bf 74} 065012 (2006)] claimed that the Landau pole of the one-loop coupling 
at $Q^2=\Lambda^2$  is absent  from  the leading one-chain term in 
a skeleton expansion of the Euclidean Adler ${\cal D}$ function. Moreover, in this
approximation  one has continuity along the Euclidean axis and a 
smooth infrared freezing,  properties known to be satisfied by the "true" Adler function.  We show that crucial in
the derivation of these results  is the use of a modified Borel summation, which leads
simultaneously to the loss of another fundamental property of the 
true Adler function: the analyticity implied by the K\"allen-Lehmann 
representation. 
\end{abstract}
\maketitle
 \end{titlepage} 
\newpage\section{Introduction}
 In confined gauge theories like QCD,  causality and unitarity imply that 
 the Green functions and the physical amplitudes are analytic functions of
  the complex energy variables, with singularities at the hadronic unitarity
   thresholds \cite{Oehme}.
In particular, the Adler function ${\cal D}(Q^2)$ (related to the polarisation 
amplitude by $ {\cal D}(Q^2)=-Q^2 \mbox{d} \Pi/\mbox{d} Q^2 -1$) is a real analytic function
in the complex $Q^2$ plane cut along the negative real axis from the 
threshold  $- 4 M_\pi^2$ for hadron production  to $-\infty$.  This property 
is implemented by the well-known K\"allen-Lehmann representation
\begin{equation}\label{KL}
 {\cal  D}(Q^2)=\frac{Q^2}{\pi} \int\limits_{4 M_\pi^2}^\infty \frac{{\cal R}(s)\, \mbox{d} s}{(s+Q^2)^2}\,,
\end{equation}
where ${\cal R}(s)$ is related to the observable cross section $\sigma_{e^+e^-\to {\rm hadrons}}$. From (\ref{KL}) it follows in particular that ${\cal  D}(Q^2)$ is continuous in the Euclidean region $Q^2>0$ and vanishes at $Q^2=0$.

The renormalization-group improved expansion of the Adler function  
in massless QCD does not satisfy all the properties contained in the above representation. 
The finite-order expansion
\begin{equation}\label{Dseries} 
{\cal D}^{(N)}_{PT} (Q^2) = \sum \limits_{n= 0}^N d_n\, a^{n+1}(Q^2)
\end{equation}
is plagued by the unphysical (Landau) pole  at $Q^2=\Lambda^2$, present in the one loop running coupling 
\begin{equation}\label{asrng}
   a(Q^2) = \frac{\alpha_s(Q^2)}{\pi}= \frac{1}{\beta_0\ln(Q^2/\Lambda^2)} \,.
\end{equation} 
A modified perturbative QCD series ("analytic perturbation theory"), which implements the  K\"allen-Lehmann representation (\ref{KL})  at each finite order,  has been proposed in \cite{Sh, ShSo}.

Beyond finite orders, the observables can be defined by a summation of the Borel
type. The Borel transform $B(u)$ of the Adler function has singularities on the 
real axis of the $u$-plane \cite{Beneke}: the ultraviolet (UV) renormalons  
along the range $u\leq -1$, and the infrared (IR) renormalons along $u \geq  2$ 
(we adopt the definition of the Borel transform used in \cite{MaNe}). While the 
Borel transform is, for a wide class of functions, uniquely determined once all 
the perturbation expansion coefficients are explicitly given, the determination 
of the function having a given perturbative (asymptotic) expansion is, actually, infinitely ambiguous; not 
only due to the singularities, but 
because the contour of the Borel-type integral can be also varied,  
without affecting the expansion coefficients of the 
perturbation series.   

  In Ref. \cite{BrMa} the authors use two different Borel summations of the perturbation series in the Euclidean region:
 for positive coupling,  
$a(Q^2)>0$, they choose the integration contour along the positive (IR renormalon) axis, 
\begin{equation}\label{Borelp}   
{\cal D}_{PT} (Q^2)=\frac{1}{\beta_0}
\int\limits_0^\infty\!\mbox{e}^{-u/(\beta_0 a(Q^2))} \, B(u)\,{\rm d}u, \quad\quad a(Q^2)>0,
\end{equation}
 while for negative coupling the integral is taken instead along the negative 
 (UV renormalon) axis:
\begin{equation}\label{Boreln}   
{\cal D}_{PT}(Q^2)=\frac{1}{\beta_0}
\int\limits_0^{-\infty}\!\mbox{e}^{-u/ (\beta_0 a(Q^2))} \, B(u)\,{\rm d}u,\quad \quad a(Q^2)<0.
\end{equation}
As shown in \cite{BrMa}, the  summation based on the above definitions can be expressed as:
\begin{equation}\label{Domega}
   {\cal D}_{PT}(Q^2) =  \int\limits_0^\infty\!   {\rm d}\tau\,\omega_{\cal D}(\tau)  a(\tau Q^2)\,, \end{equation}
  in terms of the 
characteristic function $\omega_{\cal D}(\tau)$ defined by Neubert \cite{MaNe}. Regulating  
with the Principal 
Value the singularity  of $a(\tau Q^2)$ at $\tau=\Lambda^2/Q^2$, and taking into account the continuity of the 
 characteristic function $\omega_{\cal D}(\tau)$  at $\tau=1$, the authors 
 of \cite{BrMa} conclude
  from (\ref{Domega}) that the contribution of the leading chain of the 
  skeleton expansion of  the Adler function 
 is finite and continuous along the whole spacelike  axis $Q^2>0$ and approaches a
  zero limit at $Q^2=0$.
  
Therefore, in \cite{BrMa} it is shown that  by a suitable summation of a class of 
diagrams in perturbative QCD, one recovers a  property of the true Adler function,
which follows from the representation (\ref{KL}). Unfortunately, it turns out that 
another fundamental property implied by same representation (\ref{KL}), namely 
analyticity in the complex plane, is simultaneously lost. In the present Comment,  
we prove this by calculating the Adler function in the complex energy plane with 
 the Borel prescription adopted in \cite{BrMa}.  The calculation uses the technique 
 described in \cite{CaNe}, based on the inverse Mellin transform of the Borel function. 

\section{Characteristic function and inverse Mellin transform}

 As shown in \cite{MaNe}, the function $\omega_{\cal D}$  appearing in (\ref{Domega}) is the inverse Mellin transform  of the Borel function $B(u)$:
\begin{equation}\label{omega}   
\omega_{\cal D}(\tau) = \frac{1}{2\pi i} 
  \int\limits_{u_0-i\infty}^{u_0+i\infty}\!{\rm d}u\,   B(u)\,\tau^{u-1} \,. 
  \end{equation}
The inverse relation 
\begin{equation}\label{omegainv}   
B(u) = \int\limits_0^\infty\!{\rm d}\tau\, 
  \omega_{\cal D}(\tau)\,\tau^{-u} \,, 
  \end{equation} 
  defines the function $B(u)$ in a strip parallel to the
imaginary axis with $-1<\mbox{Re}\,u <2$, where it is assumed to be analytic.  

The function $\omega_{\cal D}(\tau)$ was calculated in \cite{MaNe}  
in the large-$\beta_0$ approximation.  The result was rederived 
in  \cite{BrMa}.
 Using (\ref{omega}), the calculation is based on residues theorem: for
 $\tau<1 $ the integration contour is closed on the right half-$u$-plane, 
 and the result is the sum over the residues  of 
the infrared renormalons;   for
 $\tau>1 $ the integration contour is closed on the left half-$u$-plane, 
 and the result  contains the residues
 of the  ultraviolet renormalons.  The residues of the IR and UV renormalons
  satisfy some symmetry properties \cite{BrMa}, but their contributions are 
  not equal. Therefore  $\omega_{\cal D}(\tau)$  has different analytic 
expressions, depending on whether 
$\tau$ is less or greater than 1. 
Following Ref. \cite{BrMa}, we denote the two branches of $\omega_{\cal D}$  
by $\omega_{\cal D}^{IR}$ and  $\omega_{\cal D}^{UV}$,
 respectively. 
  According to the above discussion, it follows from (\ref{omega}) that
\begin{equation}\label{omegaIR}   
\omega_{\cal D}^{IR}(\tau) = \frac{1}{2\pi i} \left[\,
  \int\limits_{{\cal C_+}}\!{\rm d}u\,   B(u)\,\tau^{u-1}- 
\int\limits_{{\cal C_-}}\!{\rm d}u\,   B(u)\,\tau^{u-1} \right] \,, 
  \end{equation}
 where ${\cal C_\pm}$  are two parallel lines going from $0$ to $+\infty$ 
slightly above and below the real positive axis, and
\begin{equation}\label{omegaUV}   
\omega_{\cal D}^{UV}(\tau) = \frac{1}{2\pi i} \left [\,
  \int\limits_{{\cal C_+'}}\!{\rm d}u\,   B(u)\,\tau^{u-1}- 
\int\limits_{{\cal C_-'}}\!{\rm d}u\,   B(u)\,\tau^{u-1}\,\right] \,, 
  \end{equation}
where ${\cal C_\pm'}$  are two  lines  going from $0$ to $-\infty$ 
slightly above and below the real negative axis.

The explicit expressions of  $\omega_{\cal D}^{IR}$ and $\omega_{\cal D}^{UV}$ in
 the large-$\beta_0$ approximation are given in Eq. (80) of \cite{MaNe} (see also Eq. (2.19)
 of \cite{CaNe}, where $\omega_{\cal D}^{IR}$ is 
 denoted by $\widehat w_D^{(<)}$, and $\omega_{\cal D}^{UV}$ by $\widehat w_D^{(>)}$).
As shown in \cite{MaNe}, the function $\omega_{\cal D}(\tau)$  and its first 
three derivatives are continuous  at $\tau=1$.  Moreover, the explicit expressions given
 in \cite{MaNe, CaNe} imply that $\omega_{\cal D}^{IR}(\tau)$   
 and  $\omega_{\cal D}^{UV}(\tau)$  are both  analytic  functions 
  in the $\tau$-complex plane cut along the real negative axis $\tau<0$.

 \section{Adler function in the complex plane}
A closed representation of the Adler function ${\cal D}_{PT}(Q^2)$  for complex values 
of $Q^2$ in terms of the 
characteristic function   was derived in  \cite{CaNe}.
 The function  ${\cal D}_{PT}(Q^2)$  was defined  for large $|Q^2|$ by a Borel-Laplace integral
 along the IR axis,  while the expression for low $Q^2$ was obtained by analytical 
 continuation.   In the present Comment we use the technique presented in \cite{CaNe}, 
 adapted for the choice of the Borel-Laplace integral made in \cite{BrMa}. For clarity, 
 we shall present the calculation in some detail. 

As in Ref. \cite{BrMa}
we work in the $V$-scheme, where all the exponential dependence in the Borel-Laplace integrals (\ref{Borelp}) and (\ref{Boreln}) is 
absorbed in the running coupling, and denote by $\Lambda_V^2$  the corresponding QCD scale parameter.
 Let us consider $Q^2$ complex, first such that $|Q^2|>\Lambda_V^2$. Since in 
this case $\mbox{Re}\, a(Q^2)>0$ we use the choice (\ref{Borelp}) 
of the Borel-Laplace integral with the principal value ($PV$) prescription, taking
\begin{equation}\label{pv}
{\cal D}_{PT}(Q^2)={1\over 2} [{\cal D}^{(+)}(Q^2)+ {\cal D}^{(-)}(Q^2)]\,,
\end{equation}
where ${\cal D}^{(\pm)}(Q^2)$ are defined as
\begin{equation}\label{pm}   
{\cal D}^{(\pm)}(Q^2)=\frac{1} {\beta_0}
\int\limits_{{\cal C}_\pm}\!{\rm e}^{-u/(\beta_0 a(Q^2))} \, B(u)\,{\rm d}u\,.
\end{equation} Here ${\cal C_\pm}$  are two parallel lines 
slightly above and below the real positive axis, introduced already in Eq. (\ref{omegaIR}).

Following \cite{CaNe}, we pass from the integrals along the contours ${\cal C_\pm}$ 
to integrals along a line parallel to the imaginary axis, 
where the representation (\ref{omegainv}) is valid. 
This can be achieved by rotating the integration contour from the real
 to the imaginary axis, provided the contribution of the circles  at infinity
  is negligible.  We consider first  a point in the upper half of the energy plane, for which $Q^2=|Q^2|\,e^{i\phi}$ with a phase 
$0<\phi<\pi$. Taking $u={\cal R}\,e^{i\theta}$ on a large semi-circle 
of radius ${\cal R}$, the relevant exponential appearing in the 
integrals (\ref{pm}) is
\begin{equation}\label{expon}
   \exp\left\{ -{\cal R}  \left[\ln\left(\frac{|Q^2|}{\Lambda_V^2}\right)\cos\theta 
   -\phi \sin\theta \right] \right\} \,.
\end{equation}
For $|Q^2|>\Lambda_V^2$, the exponential is negligible at large ${\cal R}$ 
for $\cos\theta>0$ and $\sin\theta<0$, {\it i.e.} for the fourth quadrant of 
the complex $u$-plane. The integration contour 
defining $ {\cal D}^{(-)}(Q^2)$ can be rotated to the negative imaginary $u$-axis, where
the representation (\ref{omegainv}) is valid. This leads to the double
integral
\begin{equation}\label{double}
   {\cal D}^{(-)}(Q^2) = \frac{1} {\beta_0} \int\limits_0^{-i\infty}\!{\rm d}u
   \int\limits_0^\infty\!{\rm d}\tau\,\omega_{\cal D}(\tau) \exp\left[
   -u  \left( \ln\frac{\tau|Q^2|}{\Lambda_V^2} + i \phi \right)
   \right] \,.
\end{equation}
The order of integrations over $\tau$ and $u$ can be interchanged,
since for positive $\phi $ the integral over $u$ is convergent and can
be easily performed. Expressed  in terms of the complex variable $Q^2$, the result is
\begin{equation}\label{Dminus}
    {\cal D}^{(-)}(Q^2)  = \frac{1}{\beta_0} \int\limits_0^\infty\!{\rm d}\tau\,
   \frac{\omega_{\cal D}(\tau)}{\ln(\tau Q^2/\Lambda_V^2)}= \int\limits_0^\infty\!{\rm d}\tau\,
   \omega_{\cal D}(\tau) a(\tau Q^2) \,.
\end{equation}
We  evaluate now  the function $  {\cal D}^{(+)}(Q^2) $ given by 
the integral along the contour ${\cal C_+}$ above the real axis. The rotation
of the integration contour  to the 
positive imaginary axis is not allowed, because along the corresponding 
quarter of a circle $\sin\theta>0$, and the exponent (\ref{expon}) does not 
vanish at infinity for $0<\phi$. As explained in \cite{CaNe}, we must perform again a 
rotation to the negative imaginary $u$ axis, for which the 
contribution of the circle at infinity vanishes. But in this rotation 
the contour crosses the positive real axis, and hence picks up 
the contributions of the IR renormalon singularities located along 
this line. This can be evaluated by comparing the expression (\ref{omegaIR}) 
of the function $\omega_{\cal D}^{IR}(\tau)$ with the definition 
(\ref{pm}) of the functions ${\cal D}^{(\pm)} $: they are connected by the 
change of variable $\tau=\exp [-1/(\beta_0 a(Q^2))]$. 
It follows that ${\cal D}^{(+)} $ can 
  be expressed in terms of   ${\cal D}^{(-)}$ as
\begin{equation}\label{dif}
   {\cal D}^{(+)}  =  {\cal D}^{(-)} + \frac{2\pi i}{\beta_0} \,
   \frac{\Lambda_V^2}{Q^2}\, \omega_{\cal D}^{IR} (\Lambda_V^2/Q^2) \,.
\end{equation}
The relations (\ref{pv}), (\ref{Dminus})  and (\ref{dif}) completely specify the
function ${\cal D}_{PT}(Q^2)$ for $ |Q^2|>\Lambda_V^2$,  
in the upper half plane $\mbox{Im}\, Q^2>0$ :
\begin{equation}
 {\cal D}_{PT}(Q^2) =   \int\limits_0^\infty\!   {\rm d}\tau\,\omega_{\cal D}(\tau) a(\tau Q^2)
   + \frac{i\pi}{\beta_0}\,  \frac{\Lambda_V^2}{Q^2} \,   \omega_{\cal D}^{IR}(\Lambda_V^2/Q^2)\,.
\end{equation}
Using the same method, the function ${\cal D}_{PT}(Q^2)$ can be calculated in the 
lower half of the energy plane, where $Q^2=|Q^2|\mbox{e}^{i\phi}$ with
$-\pi<\phi<0$. In this case, the integral along ${\cal C_+}$ can be 
calculated by rotating the contour up to the positive imaginary $u$ 
axis, while for the integration along ${\cal C_-}$ one must first pass 
across the real axis and then rotate towards the positive imaginary axis. Combining the results, we obtain the following
expression for the Adler function for complex $Q^2$ with $|Q^2|>\Lambda_V^2$:
\begin{equation}\label{Dup}
   {\cal D}_{PT}(Q^2) =   \int\limits_0^\infty\!   {\rm d}\tau\,\omega_{\cal D}(\tau) a(\tau Q^2)
   \pm \frac{i\pi}{\beta_0} \, \frac{\Lambda_V^2}{Q^2} \,   \omega_{\cal D}^{IR}(\Lambda_V^2/Q^2) \,,
\end{equation}
where the $\pm$ signs correspond to $\mbox{Im}\, Q^2>0$ and $\mbox{Im}\, Q^2<0$, respectively.
We recall that the first term in (\ref{Dup}) is given by the integration with respect to $u$, 
while the last term is produced by the 
residues of the infrared renormalons picked up by crossing the positive axis of the Borel plane. 

 We consider now $|Q^2|< \Lambda_V^2$, when $\mbox{Re}\, a(Q^2)<0$. Following \cite{BrMa} we use 
 the definition  (\ref{Boreln}) of the Borel-Laplace integral along the negative axis. In this 
 case the integral is not defined due to the UV renormalons. The Principal Value prescription 
 will be given by (\ref{pv}), where the ${\cal D}^{(\pm)}$ are now 
\begin{equation}\label{pm1}   
{\cal D}^{(\pm)}(Q^2)=\frac{1}{\beta_0}
\int\limits_{{\cal C'}_\pm}\!{\rm e}^{-u/(\beta_0 a(Q^2))} \, B(u)\,{\rm d}u\,,
\end{equation}  ${\cal C'_\pm}$  being the two parallel lines 
above and below the negative $u$-axis defined in (\ref{omegaUV}).

We apply then the same techniques as above, by rotating the 
contours ${\cal C'_\pm}$ towards the imaginary axis in the $u$ plane, 
where the representation (\ref{omegainv}) is valid. If the exponential (\ref{expon}) 
decreases we can make the rotation. If not, we must first cross the real axis and
 perform the rotation. The calculations proceed exactly as before, with the difference that now one picks up the contribution of the UV renormalons, according to the relation (\ref{omegaUV}). This leads to the expression of the Adler function for $|Q^2|<\Lambda_V^2 $
\begin{equation}\label{Dlow}
   {\cal D}_{PT}(Q^2) =  \int\limits_0^\infty\!   {\rm d}\tau\,\omega_{\cal D}(\tau)  a(\tau Q^2) 
   \pm \frac{i\pi}{\beta_0}\,  \frac{\Lambda_V^2}{Q^2} \,
\omega_{\cal D}^{UV}(\Lambda_V^2/Q^2) \,,
\end{equation}
where the signs correspond to $\mbox{Im}\, Q^2>0$ and $\mbox{Im}\, Q^2<0$, respectively.

 We show now that the limit of the expressions (\ref{Dup}) 
 and (\ref{Dlow}) when $Q^2$ is approaching the 
 Euclidean axis coincides with (\ref{Domega}). Consider first that $Q^2$  tends to 
 the real positive axis 
 from above, in the region $|Q^2|>\Lambda_V^2$, when ${\cal D}_{PT}(Q^2)$ has the 
 expression (\ref{Dup}). The integrand has a pole  at $\tau=\Lambda_V^2/Q^2$.  
 Writing explicitly the real and the imaginary part of the integral we obtain, 
 for real $Q^2>\Lambda^2$:
\begin{equation}\label{Dup+}
   {\cal D}_{PT}(Q^2+i\epsilon) =  \mbox{Re}\left[ \int\limits_0^\infty\!   
   {\rm d}\tau\,\omega_{\cal D}(\tau)  a(\tau Q^2)\right] -
    \frac{i\pi}{\beta_0} \, \frac{\Lambda_V^2}{Q^2} \,  
[ (\omega_{\cal D}(\Lambda_V^2/Q^2)-\omega_{\cal D}^{IR}(\Lambda_V^2/Q^2)]\,.
\end{equation}
But for $\Lambda_V^2/Q^2<1$, the function 
 $\omega_{\cal D}$ coincides with $\omega_{\cal D}^{IR}$, so the last term in 
 (\ref{Dup+}) vanishes: the imaginary part  of the integral
  in (\ref{Dup})
 is exactly compensated by the additional term.

 For $Q^2<\Lambda_V^2$, we obtain  from (\ref{Dlow})
\begin{equation}\label{Dlow+}
   {\cal D}_{PT}(Q^2+i\epsilon)=
 \mbox{Re}\left[ \int\limits_0^\infty\!   {\rm d}\tau\,\omega_{\cal D}(\tau)  
 a(\tau Q^2)\right] -
    \frac{i\pi}{\beta_0} \, \frac{\Lambda_V^2}{Q^2}\,  
[ (\omega_{\cal D}(\Lambda_V^2/Q^2)-\omega_{\cal D}^{UV}(\Lambda_V^2/Q^2)]\,,
\end{equation}
in the same way. Again the last term in this relation vanishes, since for
$\Lambda_V^2/Q^2>1$ the function  $\omega_{\cal D}$ is equal to $\omega_{\cal 
D}^{UV}$. Moreover, one can easily see that the expressions of 
${\cal D}_{PT}(Q^2-i\epsilon)$, obtained for $Q^2$  approaching the Euclidean 
axis from the lower half plane, differ from (\ref{Dup+}) and (\ref{Dlow+}) 
only by the sign in front of the last term, which again vanishes. Thus, for 
all $Q^2>0$, the functions (\ref{Dup}) and (\ref{Dlow}) approach the same 
expression
\begin{equation}\label{D+-}
   {\cal D}_{PT}(Q^2\pm i\epsilon) =  \mbox{Re}\left[ \int\limits_0^\infty\!   {\rm d}\tau\,\omega_{\cal D}(\tau)  a(\tau Q^2)\right]\,.
\end{equation} 
This coincides with the PV regulated integral of the Cauchy type 
 (\ref{Domega}) which, as shown in \cite{BrMa}, is finite and satisfies the 
 infrared freezing. 
 Moreover, since $\omega_{\cal D}(\tau)$  is holomorphic (infinitely 
 differentiable) for all $\tau>0$ except $\tau=1$, the right-hand side of 
 (\ref{D+-}) has all derivatives defined at $Q^2>0$, except at $Q^2=\Lambda^2$,  
 where only the first three derivatives exist \cite{MaNe}. This means that 
 (\ref{Dup}) and (\ref{Dlow}) define in fact analytic functions in the regions 
 $|Q^2|>\Lambda_V^2$ and $|Q^2|<\Lambda_V^2$, respectively. In this way we have
 obtained, following the approach of Ref. \cite{BrMa}, two expressions, 
  (\ref{Dup}) and (\ref{Dlow}), which represent ${\cal D}_{PT}(Q^2)$ in
  terms of analytic functions for $|Q^2|>\Lambda_V^2$ and $|Q^2|<\Lambda_V^2$ respectively.

But the success is illusory, because 
$\omega_{\cal D}^{IR}(\tau)$ 
 and $\omega_{\cal D}^{UV}(\tau)$ are {\em two different analytic functions}. 
 The expressions (\ref{Dup}) and (\ref{Dlow}) show that ${\cal D}_{PT} (Q^2)$ 
 coincides with a certain analytic function in the region $|Q^2|>\Lambda_V^2$, 
 but with another analytic function in the region $|Q^2|<\Lambda_V^2$. So, the 
 Adler function obtained with the two different Borel representations adopted 
 in \cite{BrMa} is not analytic, but only piecewise analytic. This is in 
 evident conflict with the principle of analyticity implemented by the 
 K\"allen-Lehmann representation (\ref{KL}). 
 
The above discussion refers only to the calculation in perturbation theory. In Ref. 
\cite{BrMa}, the authors add a nonperturbative term to the perturbative Adler function. 
From  Eq. (81) of \cite{BrMa} it follows that the 
nonperturbative part added to the perturbative function ${\cal D}_{PT}(Q^2)$ given 
in our relations (\ref{Dup}) and (\ref{Dlow})  has the form 
\begin{equation}\label{DNP}
   {\cal D}_{NP}(Q^2) = \kappa \, \frac{\Lambda_V^2}{Q^2} \, 
 \omega_{\cal D}(\Lambda_V^2/Q^2)\,,
\end{equation}
where $\kappa$ is a real constant.   Using the fact that   
$\omega_{\cal D}(\Lambda_V^2/Q^2)$ behaves at small $Q^2$ like 
$Q^4/\Lambda_V^4 \ln (\Lambda_V^2/Q^2)$  \cite{MaNe}, one can see 
from  the relations (\ref{D+-}) and (\ref{DNP}) that the  sum  
${\cal D}_{PT}(Q^2)+ {\cal D}_{NP}(Q^2)$ is finite along the  Euclidean 
axis and  vanishes at $Q^2=0$. But it fails to be a single analytic 
function in the complex $Q^2$-plane, being only piecewise analytic.

\section{Discussion}

 We have shown by explicit calculation that the Borel prescription adopted 
 in \cite{BrMa} is in conflict with analyticity, which is a general property
 considered fundamental in field theory. This result implies that the infrared 
 freezing of the Euclidean observables achieved in \cite{BrMa} has had a price, 
 being possible only at the expense of analyticity. The loss is not only of an 
 academic interest: the analytical continuation is the only technique to obtain 
 the Minkowskian observables form the Euclidean ones, and all theoretical 
 predictions in field theory are based on it.   
 Moreover, the simple model for the complete Adler function proposed in 
 \cite{BrMa} cannot represent the physical observable: although it is finite 
 in the Euclidean region and exhibits infrared freezing, it is not consistent 
 with the analyticity properties implied by the K\"allen-Lehmann representation. 
 
Note also that analyticity  is repeatedly invoked by the authors  themselves 
(for instance, the term "analytical continuation" or its verbal analog are 
mentioned at least eight times in \cite{BrMa}, in particular in Sections VI 
and VII, where the Minkowskian ratio ${\cal R}$ is discussed). Analytical  
continuation is unavoidable even if a smearing procedure is used in the 
Minkowskian region.
  
 It is worth emphasizing that the result of Ref. \cite{BrMa} is not an 
 intrinsic or natural property of the leading one-chain term in the skeleton 
 expansion of QCD,  but the consequence of a specific, but questionable 
 hypothesis. A step of crucial importance in  \cite{BrMa}  is the ad-hoc 
 redefinition of the Borel integral in the region where the running coupling 
 $a(Q^2)$ becomes negative. In Ref. \cite{BrMa}, this redefinition originates 
 in a particular utilization of the function $\mbox{Ei}(z)$. The authors 
 expressed the  Borel integrals, cf. Eqs. (28) and (29) of  \cite{BrMa}, in 
 terms of $\mbox{Ei}(z)$  depending only on the ratio $z=a/z_n$, where $a$ 
 is the coupling and the $z_n$ are the positions of renormalons.
     With the conventional definition of the Principal Value of
      $\mbox{Ei}(z)$, a branch cut is located at $a>0$ and $z_n<0$, or at 
      $a<0$ and $z_n>0$. This implicitly selects a  specific form of  
      the  Borel integral: for $a>0$, it is taken along the positive, and  
      for   $a<0$, along the negative real semiaxis, respectively.
       But this definition is not the only  possibility. Note that, as pointed
  out in Ref. \cite{BrMa}  (Section VII), for $Q^{2}<\Lambda^{2}_{V}$ the
  expression (\ref{asrng}) is not the solution of the renormalization-group
  equation. We have shown that the use of these two different Borel-type 
  integrals defining one single function in two different regions is
  responsible for the loss of analyticity. 
  
 Incidentally,  the  authors of Ref. \cite{BrMa} admit   that the function 
 $ \mbox{Ei}(z)$ regulated by 
  the Principal Value does not give a reasonable result the for Minkowskian 
  observable ${\cal R}$.
   In Section VI they adjust the result by hand,   by introducing  additional 
   ad-hoc terms 
   (see Eqs. (89)-(92)  of  \cite{BrMa} and \cite{HoMa}). These ambiguous procedures are
    avoided if analyticity is preserved and analytic continuation is performed 
    in a consistent way \cite{CaFi2005}. 

\begin{acknowledgments} We acknowledge  interesting discussions with Chris 
Maxwell and thank Stan Glazek for useful comments.  This work was supported  
by the CEEX Program of Romanian ANCS under Contract Nr.2-CEx06-11-92, and by 
the Ministry of Education of the Czech Republic, Project Nr. 1P04LA211. 
\end{acknowledgments}

\end{document}